\documentclass[conference]{IEEEtran}

\usepackage{graphicx}
\usepackage{url}
\usepackage{amsmath,amssymb,amsfonts}
\usepackage[utf8]{inputenc}
\usepackage{color}

\newtheorem{algorithm}{Algorithm}

\begin{document}

\title{Adaptive Synchronization of Robotic Sensor Networks}

\author{\IEEEauthorblockN{Kasım Sinan YILDIRIM*}
\IEEEauthorblockA{*Department of Computer Engineering, Ege University\\
Üniversite caddesi 35100 Bornova, İzmir, Turkey\\
Email:sinan.yildirim@ege.edu.tr}
\and
\IEEEauthorblockN{Önder GÜRCAN\textdagger{}*}
\IEEEauthorblockA{\textdagger{}CEA, LIST,
Laboratory of Model driven\\ engineering for embedded systems\\
Point Courrier 174, Gif-sur-Yvette, F-91191 France\\
Email:onder.gurcan@cea.fr}}

\maketitle

\IEEEpeerreviewmaketitle

\begin{abstract}

The main focus of recent time synchronization research is developing power-efficient synchronization methods that meet pre-defined accuracy requirements. However, an aspect that has been often overlooked is the high dynamics of the network topology due to the mobility of the nodes.  Employing existing flooding-based and peer-to-peer synchronization methods, are networked robots still be able to adapt themselves and self-adjust their logical clocks under mobile network dynamics? In this paper, we present the application and the evaluation of the existing synchronization methods on robotic sensor networks.  We show through simulations that Adaptive Value Tracking synchronization is robust and efficient under mobility. Hence, deducing the time synchronization problem in robotic sensor networks into a \textit{dynamic value searching} problem  is preferable to existing synchronization methods in the literature.

\end{abstract}

\section{Introduction}

Time synchronization is one of the fundamental building blocks for coordinated and power-efficient operation of the networked robots. Ironically, time synchronization itself is also an energy consuming process which demands communication and information processing among the robots. This has led most research in time synchronization literature to focus on  developing the less power consuming synchronization method while meeting pre-defined accuracy requirements.

\begin{figure}
\centering
\includegraphics[width=\columnwidth]{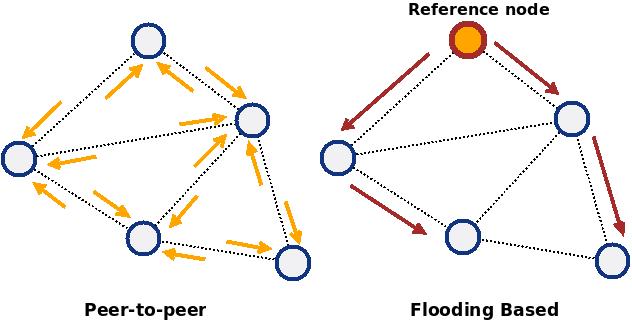}

\caption{\label{fig:comparison}In time synchronization, each robot may synchronize to the neighboring robots inside its communication range (\textit{peer-to-peer}) or to a reference robot which floods the stable time information periodically (\textit{flooding-based}).}
\end{figure}

Roughly, in the existing synchronization methods, participants collect noisy time information propagating through the network and construct a clock in software, so-called \textit{logical clock}, whose input is the value read from the unstable built-in clock and whose output is the network-wide global time. In general, the logical clock can be constructed via collecting stable time information flooded by a special reference node or via peer-to-peer communication where nodes interact with and synchronize  to their direct neighbors, as presented in Figure \ref{fig:comparison}. Least-squares \cite{Maroti:2004:FTSP,Lenzen:2009:PulseSync,Yildirim:2012:Drift-Estimation-Using-Pairwise-Slope-with-Minimum-Variance-in-Wireless-Sensor-Networks,Yildirim:2014:Time-Synchronization-Based-on-Slow-Flooding-in-Wireless-Sensor-Networks} and consensus based on distributed averaging  \cite{Schenato:2011:AverageTimeSync,Sommer:2009:GTSP,Yildirim:2014:External-Gradient-Time-Synchronization-in-Wireless-Sensor-Networks} are the common methods employed for this construction.  
Recently, we proposed a  novel technique for the construction of the logical clock: we considered the problem of synchronization  as a \textit{search process} in which each sensor node is trying to adjust the speed of its logical clock without knowing its correct value. We employed the \textit{Adaptive Value Trackers} (AVTs) \cite{Lemouzy:2011:Principles-and-Properties-of-a-MAS-Learning-Algorithm-A-Comparison-with-Standard-Learning-Algorithms-Applied-to-Implicit-Feedback-Assessment} which find and track \textit{dynamic} searched values in a given search space through successive feedbacks. We proposed a flooding based implementation \cite{Yildirim:2014:Efficient-Time-Synchronization-in-a-Wireless-Sensor-Network-by-Adaptive-Value-Tracking} and a peer-to-peer implementation \cite{Gurcan:2013:Self-Organizing-Time-Synchronization-of-Wireless-Sensor-Networks-with-Adaptive-Value-Trackers} of this approach. We observed that the synchronization performances of these approaches are similar to the existing solutions in the literature with drastically lower computation overhead, which make them more power-efficient. 

However, an aspect that has been often overlooked in the recent time synchronization studies is the high dynamics of the network topology due to the mobility of the nodes. Aforementioned studies assume periodical and almost reliable communication among the nodes. Their performances are evaluated on static and non-mobile topologies. On the other hand, the time information propagating in a mobile network is subject to more noise, collisions and packet losses. Due to mobility and neighborhood changes, nodes may instantaneously start to receive time information from badly synchronized nodes. Besides, nodes may be clustered and may form dense areas where packet collisions and losses occur frequently. What is more, there may be time durations during which nodes become disconnected from the network and do not receive time information. These points are crucial for the performance of time synchronization protocols and have not been explored yet. It is still unknown whether the existing solutions are still applicable under mobile network dynamics or not: Are networked robots still be able to adapt themselves and self-adjust their logical clocks while meeting the pre-defined synchronization performance? 

In this paper, we reveal by simulations that AVT synchronization is robust and preferable to existing synchronization methods under high mobile dynamics. Completely blind execution without keeping track of any neighboring node makes it particularly suitable for robotic sensor networks. As a remark, we observed that the performance of synchronization via flooding is better than the peer-to-peer approach in mobile environments. 

\section{Method}

We propose a time syncronization method using AVTs. An AVT finds and tracks a \textit{dynamic} searched value, that may change in the time due to the dynamics of the system, in a given search space as fast as possible \cite{Lemouzy:2011:Principles-and-Properties-of-a-MAS-Learning-Algorithm-A-Comparison-with-Standard-Learning-Algorithms-Applied-to-Implicit-Feedback-Assessment}. The tracking is established via the successive feedbacks coming from the \textit{owner} of the AVT (i.e. the robot) that indicate the direction that \textit{probably} lead to the searched value.
This decision is not trivial and is made by taking into account the goal of the robot.

\begin{figure}
\centering
\includegraphics[width=\columnwidth]{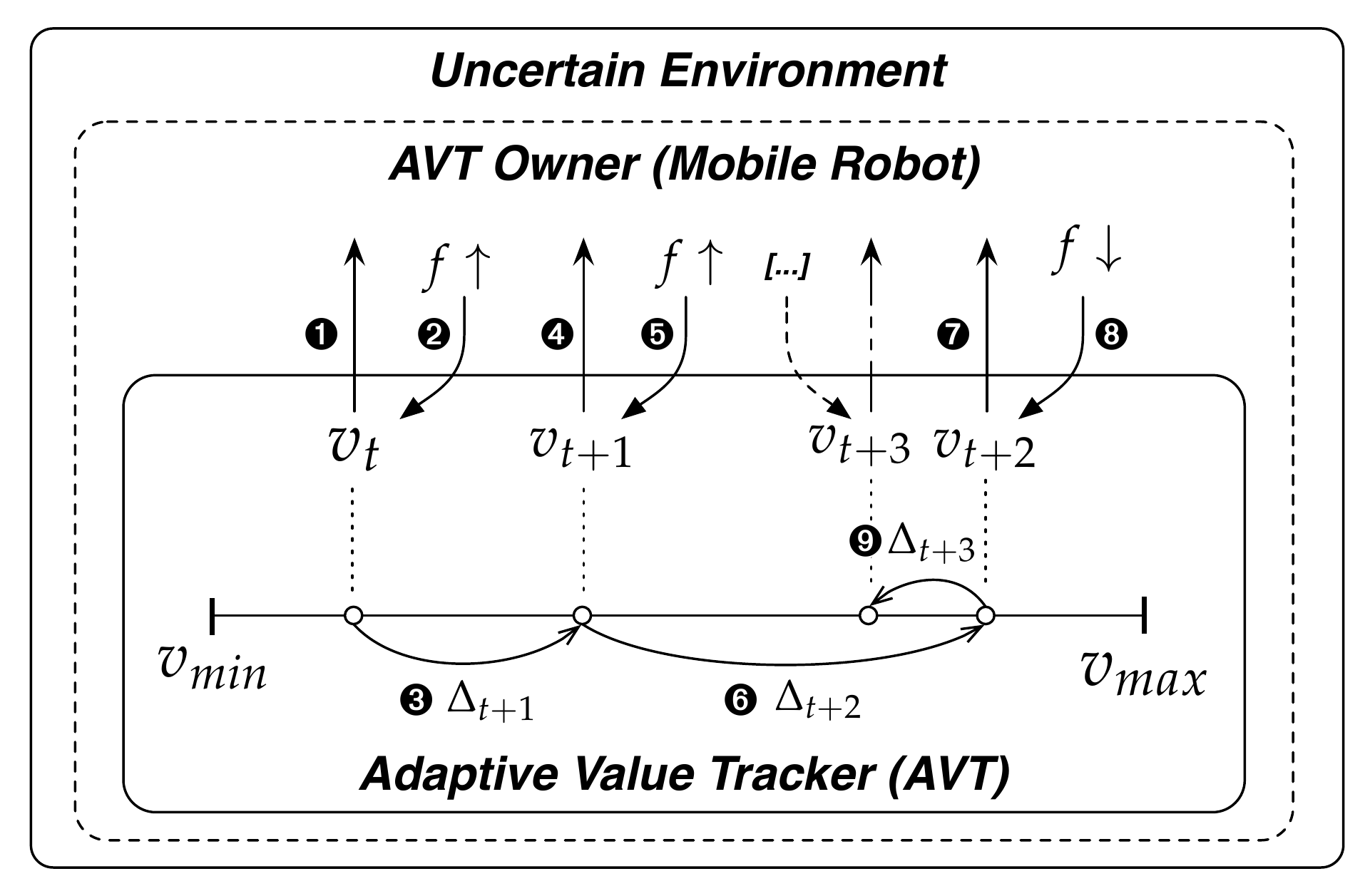}

\caption{\label{fig:avt_and_environment}Interaction between a mobile robot that situated in an uncertain environment and its AVT. The AVT value tracking process starts with an initial value $v_{0}$ and includes several cycles of search iteration until $v^{\star}$ is reached.}
\end{figure}

Formally speaking, an $avt$ searches and tracks a \textit{dynamic} value $v^{\star}$ inside a given real interval (search space) $AVT_{ss}=[v_{min},v_{max}]\subset\mathbb{R}$
where $v_{min}$ is the lower boundary and $v_{max}$ is the upper boundary for $v^{\star}$. At any time instant $t$, $avt$ is able to propose a value $v_{t}\in AVT_{ss}$ to its owner robot that can be accessed using an action of the form $v_{t}=avt.value(t)$. The objective of the robot is to determine if the
searched value $v^{\star}$ is smaller than, equal to or greater than the current proposed value $v_{t}$, without knowing the value $v^{\star}$. After this determination, the robot interacts with $avt$ using an action of the form $avt.adjust(f_{t}\in\mathcal{F})$ for sending
a \textit{feedback} $f_{t}$ from the feedback set $\mathcal{F}=\left\{ \uparrow,\downarrow,\approx\right\} $.
The feedback $f_{t}$ can be about increasing $v_{t}$ ($\uparrow$), decreasing $v_{t}$ ($\downarrow$) or informing that $v_{t}$ is good ($\approx$). A sample search with AVT is presented in Figure \ref{fig:avt_and_environment}. The details for the AVT parameters can be found in \cite{Gurcan:2013:Self-Organizing-Time-Synchronization-of-Wireless-Sensor-Networks-with-Adaptive-Value-Trackers,Yildirim:2014:Efficient-Time-Synchronization-in-a-Wireless-Sensor-Network-by-Adaptive-Value-Tracking}

In time synchronization with AVTs, each node collects periodically the time information flooded by a reference node as in \cite{Yildirim:2014:Efficient-Time-Synchronization-in-a-Wireless-Sensor-Network-by-Adaptive-Value-Tracking} or the time information of its neighboring nodes in a peer-to-peer manner as in \cite{Gurcan:2013:Self-Organizing-Time-Synchronization-of-Wireless-Sensor-Networks-with-Adaptive-Value-Trackers}. Whenever a fresh time information is received, the synchronization error is calculated by considering the value of the logical clock. In flooding based approach, the whole error is added while in the peer-to-peer approach half of the error is added to the logical clock to compensate for the clock offset. After offset compensation, each sensor node tries to find the correct speed of the logical clock with respect to its built-in clock without knowing the correct value.

%\begin{figure}
%\centering
%\includegraphics[scale=0.35]{comparison}
%\caption{\label{fig:comparison}Flooding based and peer-to-peer time synchronization approaches.}
%\end{figure}

\begin{algorithm} Speed tracking code for robot $u$
\label{alg:tracking}

$\;$1: \textbf{if} $error>0$ \textbf{then} $ avt_{u}.adjust(f\uparrow)$

$\;$2: \textbf{else if} $error < 0$  \textbf{then} $ avt_{u}.adjust(f\downarrow)$

$\;$3: \textbf{else} $ avt_{u}.adjust(f\approx)$

\end{algorithm}

A positive error indicates that the logical clock of the robot is progressing at a slower speed than the sender's logical clock. Then, a feedback about increasing the speed of the logical clock is sent to the $avt$ of that robot (Algorithm \ref{alg:tracking}, line 1). In contrast, a negative error indicates that the logical clock of the robot is progressing at a faster speed and hence a feedback about decreasing its speed is sent to the $avt$ (Algorithm \ref{alg:tracking}, line 2). Otherwise, i.e. the error is zero, $avt$ is informed that the speed of the logical clock is good (Algorithm \ref{alg:tracking}, line 3), hence it remains unchanged. 

\section{Simulation Results}

In order to evaluate the performance of time synchronization with AVT in mobile environments, we applied the flooding based and peer-to-peer protocols to the mobile robotic sensors using simulations in our discrete event simulator. In this simulator, we implemented a probabilistic radio model (Gaussian wireless channel) and a CSMA based MAC layer. Briefly, the messages are corrupted when two or more neighboring robots are trying to transmit simultaneously and messages are lost with a small probability. For mobility, we have chosen to implement random waypoint mobility model: a robot moves on a straight line to a randomly selected position in the deployment field. Once arrived, it waits for a random amount of time before it selects a new position to move to. We implemented 1 MHz built-in clocks with constant drift clock model where the drift is uniformly distributed within the interval of $\pm$ 100 parts per million. For performance comparison, we considered two other popular time synchronization protocols: PulseSync \cite{Lenzen:2009:PulseSync} and Gradient Time Synchronization Protocol (GTSP) \cite{Sommer:2009:GTSP}. PulseSync offers the time information of the reference node to be propagated as fast and reliably as possible through pulses. Receiver nodes performs least-squares regression on the received time information to construct their logical clock. On the other hand, GTSP is a peer-to-peer synchronization protocol and it performs distributed averaging for time synchronization by keeping track of the neighboring nodes.

\begin{figure*}

\centering

\includegraphics[scale=0.85]{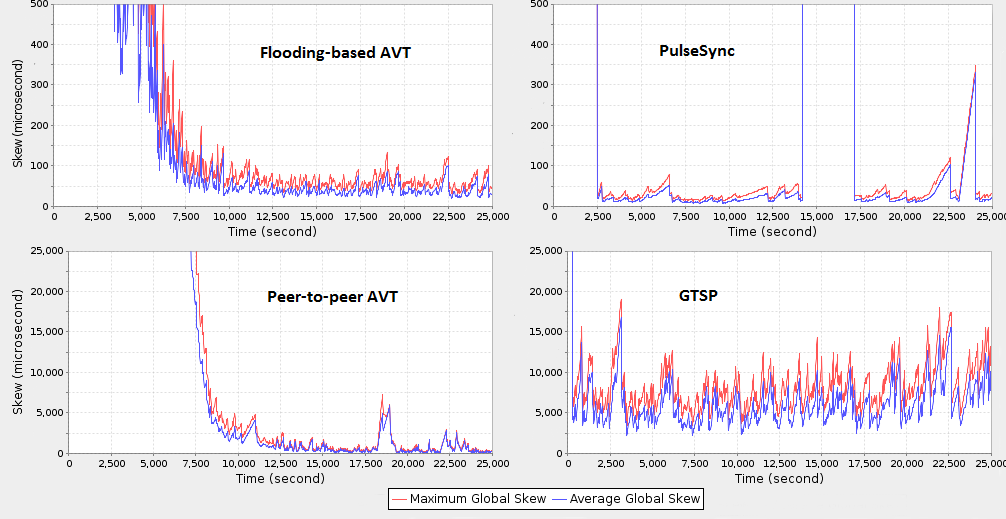}

\caption{\label{fig:simulations} Global synchronization errors observed in our simulations.}
\end{figure*}

Our evaluation metrics were the instantaneous and the average instantaneous global synchronization error: the maximum error observed between arbitrary nodes. Each of our simulation runs simulated an execution of 25000 seconds. We deployed nodes to a 300x300 meter square area randomly. The transmission range of the nodes were adjusted to 25 meters. The evaluated protocols had an identical beacon period of 30 seconds. Since the hardware clocks of sensor nodes are reported to have a drift of $\pm$ 100 parts per million, we defined the upper bound and lower bounds of the searched logical clock speed as $v_{min}=-10^{-4}$ and $v_{max}=10^{-4}$ for AVT.\footnote{Please refer to \cite{Gurcan:2013:Self-Organizing-Time-Synchronization-of-Wireless-Sensor-Networks-with-Adaptive-Value-Trackers,Yildirim:2014:Efficient-Time-Synchronization-in-a-Wireless-Sensor-Network-by-Adaptive-Value-Tracking} for the details of the parameters chosen for the successful operation of AVT.} For GTSP and PulseSync, the least-squares regression tables are composed of 8 entries. Finally, in GTSP, each node tracks at most 10 neighbors and if does not receive message during five beacon periods, it drops the information of that neighbor from the neighbor table.

In contrast to the unmobile networks, the time information propagated in a mobile network is subject to more noise, collisions and packet losses. First, while a robot is receiving a more correct and recent time information from one of its neighbors, it may not receive future information from that node due to mobility and neighborhood change. In this case, it may start to receive time information from a new neighbor whose logical clock is badly synchronized. Secondly, robots may be clustered and form dense areas where packet collisions and losses occur frequently. Last, there may be time durations during which a robot becomes disconnected from the network and may not receive time information. We observed that these points are crucial for the performance of time synchronization protocols.

Figure \ref{fig:simulations} presents the global synchronization error observed during the simulations of flooding based and peer-to-peer versions of AVT synchronization, PulseSync and GTSP. In PulseSync, the time information of the reference robot is propagated as fast as possible which reduces the noise of the received time information and decreases the time required for network-wide synchronization \cite{Lenzen:2009:PulseSync}. Flooding-based time synchronization with AVT required more time to catch the performance of PulseSync, however it is more robust to packet losses and network disconnections. In PulseSync, all of the network should be connected at pulse times of the reference robot in order to receive fresh time information. During the time interval [13000,17000] in our simulations, some robots became either disconnected from the network or too many packet losses occur in the network, hence they could not receive pulses from the reference robot. This has led these robots to loose synchronization. In contrast, in AVT synchronization, the network do not require to be connected at pulse times. Instead, each robot waits until their broadcast timer to expire in order to propagate the fresh time information of the reference robot. Any robot receiving a massage carrying a higher sequence number updates its logical clock although it might have missed the pulse of the reference robot. This makes our approach more suitable for mobile robotic networks. 

Considering peer-to-peer approaches, GTSP has crucial disadvantages compared to peer-to-peer time synchronization with AVT. First, GTSP requires robots to keep track of their neighboring robots in order to employ distributed averaging. However, it suffers from the problem of deciding which neighbors to keep track and which ones to discard in dense areas of the network, since it is not feasible to store information for all of the neighbors. Moreover, detection of the neighborhood change is another crucial problem. For instance, as a simple strategy, when a robot does not receive messages during $n$ broadcast periods from the neighbor it is currently keeping track of, it may delete its information from its internal tables. However, under high mobility, this strategy exhibits poor performance. From our simulations, we realized that GTSP is not suitable for mobile robotic networks and exhibits a poor performance. On the other hand, peer-to-peer AVT synchronization does not require to keep track of the information of the neighboring robots and it works in a \textit{completely blind} manner. Robots update their time information whenever they receive a message from their neighboring robots regardless of the identity of the sender. This strategy achieved better performance than GTSP, but not better than the flooding-based approaches.

\section{Discussion and Conclusion}

This paper presented the application and the evaluation of the recent flooding-based and peer-to-peer time synchronization methods on robotic sensor networks. It has already been shown that AVT synchronization is simple, easy to implement, memory and CPU efficient, and it establishes synchronization in finite amount of time. \cite{Yildirim:2014:Efficient-Time-Synchronization-in-a-Wireless-Sensor-Network-by-Adaptive-Value-Tracking,Gurcan:2013:Self-Organizing-Time-Synchronization-of-Wireless-Sensor-Networks-with-Adaptive-Value-Trackers}. Here in this study, we observed through simulations that AVT synchronization is also robust and efficient under mobility. 
A more detailed discussion about the robustness of AVTs can be found in \cite{Gurcan:2013:Self-Organizing-Time-Synchronization-of-Wireless-Sensor-Networks-with-Adaptive-Value-Trackers} and another discussion about the efficiency of AVTs can ben found in \cite{Yildirim:2014:Efficient-Time-Synchronization-in-a-Wireless-Sensor-Network-by-Adaptive-Value-Tracking}.
Hence, deducing the time synchronization problem in robotic sensor networks into a \textit{dynamic value searching} problem  is preferable to existing synchronization methods in the literature.

In general, the peer-to-peer approaches are expected to have a better performance in mobile networks. Intuitively, when the robots of the connected components synchronize to themselves, it should become easier to synchronize different connected components of the network. However, we observed that the flooding-based AVT synchronization performs better and establishes network-wide synchronization faster, compared to peer-to-peer strategy. We think that this mainly due to the random waypoint mobility model we applied in our simulation experiments. 
We believe that if robots apply another mobility model (where they can form group), the success of the peer-to-peer synchronization would improve. However, we leave it as a future work for now.

Besides, as another future work, we plan to explore a hybrid synchronization mechanism in which the flooding and the peer-to-peer strategies are employed together, as in \cite{Yildirim:2014:External-Gradient-Time-Synchronization-in-Wireless-Sensor-Networks}. 

\bibliographystyle{IEEEtran}
\bibliography{references}

\begin{thebibliography}{10}
\providecommand{\url}[1]{#1}
\csname url@samestyle\endcsname
\providecommand{\newblock}{\relax}
\providecommand{\bibinfo}[2]{#2}
\providecommand{\BIBentrySTDinterwordspacing}{\spaceskip=0pt\relax}
\providecommand{\BIBentryALTinterwordstretchfactor}{4}
\providecommand{\BIBentryALTinterwordspacing}{\spaceskip=\fontdimen2\font plus
\BIBentryALTinterwordstretchfactor\fontdimen3\font minus
  \fontdimen4\font\relax}
\providecommand{\BIBforeignlanguage}[2]{{%
\expandafter\ifx\csname l@#1\endcsname\relax
\typeout{** WARNING: IEEEtran.bst: No hyphenation pattern has been}%
\typeout{** loaded for the language `#1'. Using the pattern for}%
\typeout{** the default language instead.}%
\else
\language=\csname l@#1\endcsname
\fi
#2}}
\providecommand{\BIBdecl}{\relax}
\BIBdecl

\bibitem{Maroti:2004:FTSP}
\BIBentryALTinterwordspacing
M.~Mar\'{o}ti, B.~Kusy, G.~Simon, and A.~L{\'e}deczi, ``The flooding time
  synchronization protocol,'' in \emph{Proc. of the 2nd International
  Conference on Embedded Networked Sensor Systems}, ser. SenSys '04.\hskip 1em
  plus 0.5em minus 0.4em\relax New York, NY, USA: ACM, 2004, pp. 39--49.
  [Online]. Available: \url{http://doi.acm.org/10.1145/1031495.1031501}
\BIBentrySTDinterwordspacing

\bibitem{Lenzen:2009:PulseSync}
\BIBentryALTinterwordspacing
C.~Lenzen, P.~Sommer, and R.~Wattenhofer, ``Optimal clock synchronization in
  networks,'' in \emph{Proceedings of the 7th ACM Conference on Embedded
  Networked Sensor Systems}, ser. SenSys '09.\hskip 1em plus 0.5em minus
  0.4em\relax New York, NY, USA: ACM, 2009, pp. 225--238. [Online]. Available:
  \url{http://doi.acm.org/10.1145/1644038.1644061}
\BIBentrySTDinterwordspacing

\bibitem{Yildirim:2012:Drift-Estimation-Using-Pairwise-Slope-with-Minimum-Variance-in-Wireless-Sensor-Networks}
K.~S. Yildirim and A.~Kantarci, ``Drift estimation using pairwise slope with
  minimum variance in wireless sensor networks,'' \emph{Ad Hoc Netw.}, vol.~11,
  no.~3, pp. 765--777, May 2013.

\bibitem{Yildirim:2014:Time-Synchronization-Based-on-Slow-Flooding-in-Wireless-Sensor-Networks}
K.~Yildirim and A.~Kantarci, ``Time synchronization based on slow-flooding in
  wireless sensor networks,'' \emph{Parallel and Distributed Systems, IEEE
  Transactions on}, vol.~25, no.~1, pp. 244--253, 2014.

\bibitem{Schenato:2011:AverageTimeSync}
\BIBentryALTinterwordspacing
L.~Schenato and F.~Fiorentin, ``Average timesynch: A consensus-based protocol
  for clock synchronization in wireless sensor networks,'' \emph{Automatica},
  vol.~47, no.~9, pp. 1878--1886, Sep. 2011. [Online]. Available:
  \url{http://dx.doi.org/10.1016/j.automatica.2011.06.012}
\BIBentrySTDinterwordspacing

\bibitem{Sommer:2009:GTSP}
\BIBentryALTinterwordspacing
P.~Sommer and R.~Wattenhofer, ``Gradient clock synchronization in wireless
  sensor networks,'' in \emph{Proceedings of the 2009 International Conference
  on Information Processing in Sensor Networks}, ser. IPSN '09.\hskip 1em plus
  0.5em minus 0.4em\relax Washington, DC, USA: IEEE Computer Society, 2009, pp.
  37--48. [Online]. Available:
  \url{http://dl.acm.org/citation.cfm?id=1602165.1602171}
\BIBentrySTDinterwordspacing

\bibitem{Yildirim:2014:External-Gradient-Time-Synchronization-in-Wireless-Sensor-Networks}
K.~S. Yildirim and A.~Kantarci, ``External gradient time synchronization in
  wireless sensor networks,'' \emph{Parallel and Distributed Systems, IEEE
  Transactions on}, vol.~25, no.~3, pp. 633--641, March 2014.

\bibitem{Lemouzy:2011:Principles-and-Properties-of-a-MAS-Learning-Algorithm-A-Comparison-with-Standard-Learning-Algorithms-Applied-to-Implicit-Feedback-Assessment}
S.~Lemouzy, V.~Camps, and P.~Glize, ``Principles and properties of a mas
  learning algorithm: A comparison with standard learning algorithms applied to
  implicit feedback assessment,'' in \emph{Proc. of the 2011 IEEE/WIC/ACM Int.
  Conf. on Web Intelligence and Intelligent Agent Technology - Vol. 02}, ser.
  WI-IAT '11.\hskip 1em plus 0.5em minus 0.4em\relax Washington, DC, USA: IEEE
  Computer Society, 2011, pp. 228--235.

\bibitem{Yildirim:2014:Efficient-Time-Synchronization-in-a-Wireless-Sensor-Network-by-Adaptive-Value-Tracking}
K.~S. Yildirim and {\"O}.~G{\"u}rcan, ``Efficient time synchronization in a
  wireless sensor network by adaptive value tracking,'' \emph{Wireless
  Communications, IEEE Tran. on}, to appear.

\bibitem{Gurcan:2013:Self-Organizing-Time-Synchronization-of-Wireless-Sensor-Networks-with-Adaptive-Value-Trackers}
{\"O}.~G{\"u}rcan and K.~S. Yildirim, ``Self-organizing time synchronization of
  wireless sensor networks with adaptive value trackers,'' in
  \emph{Self-Adaptive and Self-Organizing Systems (SASO), 2013 IEEE Sixth Int.
  Conf. on}, sept. 2013, pp. 91 --100.

\end{thebibliography}

\end{document}